\documentclass[twocolumn,superscriptaddress,footinbib,bibnotes,floatfix,longbibliography]{revtex4-2}
\usepackage{amsmath}
\usepackage{amssymb}
\usepackage{graphicx}
\usepackage{bm}
\usepackage{dcolumn}
\usepackage{graphics}
\usepackage{array}

\DeclareMathOperator{\arccosh}{arccosh}
\begin{document}

\title{Superfluid helium film may greatly increase the storage time of ultracold neutrons in material traps}
\author{P. D. Grigoriev}
\affiliation{L. D. Landau Institute for Theoretical Physics, Chernogolovka,
	Moscow region, 142432, Russia} 
\affiliation{National University of Science
	and Technology MISiS, 119049, Moscow, Russia}
\email{grigorev@itp.ac.ru}
\author{A. M. Dyugaev}
\affiliation{L. D. Landau Institute for Theoretical Physics, Chernogolovka,
	Moscow region, 142432, Russia}

\begin{abstract}
We propose a method to increase both the neutron storage time and the precision of its lifetime measurements
by at least tenfold. The storage of ultracold neutrons (UCN) in material traps now provides the most
accurate measurements of neutron lifetime and is used in many other
experiments. 
The precision of these measurements is limited by the
interaction of UCN with the trap walls. We show that covering of trap
walls with liquid helium may strongly decrease the UCN losses from
material traps. $^4$He does not absorb neutrons at all. Superfluid
He covers the trap walls as a thin film, $\sim 10$ nm thick, due to the 
van der Waals attraction. 
However, this He film on a flat wall is too thin to protect the UCN from their 
absorption inside a trap material.
By combining the van der Waals attraction with capillary effects we show 
that surface roughness may increase the thickness of this film much beyond
the neutron penetration depth $\sim 33$nm. 
Using liquid He for UCN storage requires low temperature $T<0.5$ to avoid 
neutron interaction with He vapor, while the neutron losses
because of the interaction with surface waves are small and can be accounted
for using their linear temperature dependence.
\end{abstract}

\date{\today }
\maketitle

\section{Introduction}

Slow neutrons play an important role in particle physics, both as a tool and
an object. The properties of a free neutron and its interactions with known
or hypothetic fields provide a valuable information about fundamental
particles and interactions.\cite%
{Abele/2008,Musolf/2008,Dubbers/2011,WietfeldtColloquiumRMP2011,GONZALEZALONSO2019165}
The precise measurements of neutron lifetime $\tau _{n}$ are important for
elementary particle physics, astrophysics and cosmology (see \cite%
{Abele/2008,Musolf/2008,Dubbers/2011,WietfeldtColloquiumRMP2011,GONZALEZALONSO2019165}
for reviews). The search for a non-vanishing electric dipole moment of
neutrons\cite{Pospelov2005,Baker/2006,SerebrovJETPLetters2014} impose the
limits on CP violation. Precise measurements of a $\beta $-decay asymmetry
provide information on axial-vector weak coupling constant \cite%
{Beam2019PhysRevLett.122.242501,PhysRevLett.105.181803,PhysRevC.101.035503}.
The resonant transions between discrete quantum energy levels of neutrons in
the earth gravitational field\cite%
{NesvizhevskyNature2002,UCNResonancePhysRevLett.112.151105} probe the
gravitational field on a micron length scale and impose constraints on dark
matter.

A large class of experiments employs neutrons with energy lower than the
neutron optical potential of typical materials, i.e.\ $\lesssim 300$ neV 
\cite%
{Golub/1991,Ignatovich/1990,Ignatovich1996,PhysRevLett.105.181803,PhysRevC.101.035503,Baker/2006,SerebrovJETPLetters2014,NesvizhevskyNature2002,Serebrov2008PhysRevC.78.035505,ArzumanovPhysLettB2015,Serebrov2017,Serebrov2018PhysRevC.97.055503,Review2019Pattie}%
. These so-called ultracold neutrons (UCNs) can be trapped for many minutes
in well-designed "neutron bottles"\cite%
{Serebrov2008PhysRevC.78.035505,ArzumanovPhysLettB2015,Serebrov2017,Serebrov2018PhysRevC.97.055503,Review2019Pattie}%
. The gravitational interaction with a potential difference of $100$ neV per
meter rise plays important role in UCN storage and manipulation\cite%
{Golub/1991,Ignatovich/1990,Ignatovich1996,Serebrov2008PhysRevC.78.035505,ArzumanovPhysLettB2015,Serebrov2017,Serebrov2018PhysRevC.97.055503}%
. Because of of the neutron magnetic moment of $60$ neV/T,
magneto-gravitational trapping of UCN is feasible too\cite%
{Huffman2000,PhysRevC.94.045502,PhysRevC.95.035502,Ezhov2018,Pattie2018}.

The main alternative to using UCN in neutron lifetime measurements is the
cold neutron beam\cite%
{BeamPhysRevC.71.055502,BeamPhysRevLett.111.222501,BeamReview2020}, giving $%
\tau _{n}=(887.7\pm 1.2$[stat]$\pm 1.9$[syst]$)$s. Using UCN in $\tau _{n}$
measurements is believed to be much more precise. Therefore, according to
the Particle Data Group, the conventional neutron lifetime value $\tau
_{n}=879.4\pm 0.6$ s is only based on the UCN experiments \cite%
{Serebrov2008PhysRevC.78.035505,ArzumanovPhysLettB2015,Serebrov2018PhysRevC.97.055503,Ezhov2018,Pattie2018}%
. The discrepancy between these two methods is far beyond the estimated
error. This \textquotedblleft neutron lifetime puzzle\textquotedblright\ is
a subject of extensive discussion till now\cite%
{BeamReview2020,DarkMatter2021PhysRevD.103.035014,Serebrov_2019,Serebrov2021PhysRevD.103.074010}%
. Presumably, it is due to unconsidered systematic errors in beam experiments%
\cite{Serebrov2021PhysRevD.103.074010}, but their origin is not understood
yet. Other possibilities involving new physics, such as additional neutron
decay channels or dark matter \cite%
{BeamReview2020,DarkMatter2021PhysRevD.103.035014}, most probably, 
are not the reason for this discrepancy, as was shown by analyzing the 
neutron $\beta$-decay asymmetry \cite{Dubbers2018}. The
accuracy of current UCN lifetime measurements also requires additional
analysis, because unconsidered or underestimated systematic errors in
UCN experiments are also possible.

The main problem with the UCN is their storage. The traditional materials
for UCN trap walls are\cite{Ignatovich/1990} beryllium, beryllium oxide,
nickel, diamond-like carbon, copper, aluminium and others. They have low
neutron loss coefficient $\eta $ and high potential barrier $V_{0}$ for UCN.
For beryllium the theoretical value of loss coefficient $\eta $ $\sim
10^{-6}-10^{-5}$ depending on temperature\cite{Golub/1991} and the potential
barrier $V_{0}^{Be}=252\ \mathrm{neV}$. The corresponding neutron
penetration depth into Be is $\kappa _{0Be}^{-1}=\hbar /\sqrt{2mV_{0}^{Be}}%
\approx 9\ \mathrm{nm}$, where the neutron mass $m=1.675\times 10^{-24}$g.
The first experiments with UCN material traps were discouraging and
demonstrated too short neutron storage time of few minutes.\cite%
{Golub/1991,Ignatovich/1990} The origin of such strong neutron losses was
puzzling for more than a decade. Finally these losses were shown to
originate mainly from inelastic neutron scattering because of surface
contamination by hydrogen\cite{Ageron1985} (see also \cite%
{Golub/1991,Ignatovich/1990} for a review). Various methods were applied to
reduce the surface contamination.\cite{Golub/1991,Ignatovich/1990} For
example, the material traps were equipped with sputtering heads, enabling
fresh surfaces on the walls which had never been exposed to the atmosphere.
This improvement finaly reduced the systematic error in $\tau _{n}$ to $\sim
10$s in Be or Al traps with surface covered with heavy watter or oxigen.\cite%
{Kosvintsev1986,MOROZOV1989,KHARITONOV1989}

A new important step was made by using Fomblin oil or grease to cover
the UCN trap walls \cite%
{MAMBOPhysRevLett.63.593,Serebrov2008PhysRevC.78.035505,PICHLMAIER2010,Serebrov2017,ArzumanovPhysLettB2015,Serebrov2018PhysRevC.97.055503}%
. It has a pseudo Fermi potential for UCN of $V_{0}^{F}=106$ neV and, being
hydrogen free, its loss probability per UCN wall collision is $\approx
10^{-5}$ at room temperature below the potential threshold.

The precision of neutron lifetime measurements is determined by the accuracy
of the estimate of neutron escape rate from the traps, which is the main
source of systematic errors.\cite%
{Golub/1991,Ignatovich/1990,Goremychkin2017,Serebrov2018PhysRevC.97.055503,Review2019Pattie}
This estimate is based on extrapolation of the measured lifetime $\tau _{1}$
of neutrons stored in the trap to the zero neutron losses by a careful
variation of the bottle geometry and/or temperature so that the wall loss
contribution can be accurately determined. The highest precision of $\tau
_{n}$ measurements, with the uncertainty of only $\delta \tau _{n}\sim 1$s,
was announced in large Fomblin-coated material traps \cite%
{Serebrov2008PhysRevC.78.035505,Serebrov2017,Serebrov2018PhysRevC.97.055503}
and in magneto-gravitational UCN traps\cite{Pattie2018}. The corresponding
neutron lifetime values vary from $\tau _{n}=877.7$s \cite{Pattie2018} to $%
\tau _{n}=881.5$s \cite{Serebrov2017,Serebrov2018PhysRevC.97.055503}. Such a
high precision is achieved by using a large gravitational trap covered with
Fomblin grease at low temperature $T<90K$ to reduce the inelastic neutron
scattering. The resulting UCN losses due to the interaction with trap walls
were estimated\cite%
{Serebrov2008PhysRevC.78.035505,Serebrov2018PhysRevC.97.055503} to be $\approx
1/60$ of the neutron $\beta $-decay rate. Hence, the corresponding range of
extrapolation to account for these losses was only $\sim 15$ seconds. It is
complicated to further notably increase the precision of $\tau _{n}$\
measurements without reducing the extrapolation interval. Hence, one needs
to reduce the neutron losses due to interaction with the trap walls.

A possible new step to further reduce the neutron escape rate from the trap
is to cover the trap walls with liquid $^{4}$He. $^{4}$He does not absorb
neutrons at all, but it provides a very small optical potential barrier $%
V_{0}^{He}=18.5\ \mathrm{neV}$ for the neutrons. Hence, only UCN with
kinetic energy $E<V_{0}^{He}$ can be effectively stored in such a trap. The
factor $\nu _{F}\equiv V_{0}^{F}/V_{0}^{He}\approx 5.73$ reduces the neutron
phase volume and, hence, the neutron density in the He trap $\nu
^{3/2}\approx 13.7$ times as compared to the Fomblin coating. However, the
neutron phase-volume density increases with the development of technology%
\cite{PhysRevC.99.025503,ZimmerPhysRevC.93.035503}, and this neutron density
reduction factor could become less important than the decrease of neutron
loss rate, at least in some experiments.

$^{4}$He is superfluid below $T_{\lambda }=2.17$K and covers not only the
floor but also the walls and the ceiling of the trap because of the
van-der-Waals attraction. On vertical walls the thickness of the helium film
depends on the height above the level of liquid helium, as discussed in Sec.
III below. On flat vertical walls a few centimeters above the He level and on
the ceiling of the trap the thickness of the helium film is expected to be
only $d_{He}^{\min }\approx 10$nm$<\kappa _{0}^{-1}$, while the neutron
penetration depth into the liquid helium is $\kappa _{0He}^{-1}=\hbar /\sqrt{%
2m_{n}V_{0}^{He}}\approx 33.3\ \mathrm{nm}>d_{He}$. Hence, the corresponding
tunneling exponent, approximately giving the reduction of the neutron wave
function $\psi $ on the trap wall surface due to its covering by He film, 
\begin{equation}
\psi \left( 0\right) /\psi \left( d_{He}\right) \sim \exp \left( -\kappa
_{0He}d_{He}\right) ,  \label{psiS1}
\end{equation}%
is of the order of unity: for $d_{He}=d_{He}^{\min }\approx 10$nm, $\psi
\left( 0\right) /\psi \left( d_{He}\right) \approx 0.74$. This is not
sufficient to strongly reduce the neutron losses on the trap walls and to
compensate for the decrease of neutron density.

To increase the thickness of He film on the trap walls the idea of neutron
storage in a rotating He vessel was proposed\cite%
{Bokun/1984,Alfimenkov/2009}, but the rotating liquid generates additional
bulk and surface excitations. This lead to an enhanced neutron scattering
rate, which is very complicated for estimates. Moreover, a moving surface
leads to a considerable "upscattering" of neutron, i.e. to gradual increase
of their kinetic energy, finally exceeding the potential barrier $%
V_{0}^{He}$. Therefore, one needs a time-independent covering of the trap
walls with liquid $^{4}$He.

One can use liquid He to cover only the bottom of neutron trap, where
the He film can be made arbitrarily thick. Earth's gravity prevents UCNs from
leaving a sufficiently high trap through the upper edge. Neutron escape
through the side walls can be reduced by using a very wide trap or by
side protection using a magnetic field \cite{ZimmerPhysRevC.92.015501}.
However, this partial solution of the problem of small He film thickness does
not give a sufficient advantage to use this method in current $\tau _{n}$
measurements.

Liquid He introduces new inelastic scattering mechanism for neutrons because
of their interaction with He vapor atoms and with soft thermal excitations -
the quanta of surface waves, called ripplons. The corresponding scattering
rates for a neutron on the lowest vertical level were studied recently.\cite%
{PhysRevC.94.025504} The concentration of $^{4}$He vapor exponentially
decreases with temperature, $n_{V}\propto \exp \left( -7.17/T\left[ K\right]
\right) $, and can be disregarded below $0.5K$. However, the neutron
scattering rate $w_{R}$ on ripplons depends linearly on temperature\cite%
{PhysRevC.94.025504} and, formally, cannot be discarded even below $0.5K$.
However, the amplitude of neutron-ripplon scattering is small even from the
lowest neutron energy level, so that the corresponding scattering time $%
1/w_{R}$ exceeds many hours at $T<0.5K$.\cite{PhysRevC.94.025504}\ Our
preliminary results show that the amplitude of neutron-ripplon scattering
from higher levels is even smaller. Moreover, the linear temperature
dependence of neutron-ripplon scattering rate $w_{R}\left( T\right) $ allows
its effective extrapolation to zero temperature. Thus, the problem of an
additional neutron scattering on a liquid He surface can be solved. However,
a small He film thickness remains an obstacle of using liquid He in neutron
traps.

In this paper we reanalyze the advantages and drawbacks of covering the UCN
trap by helium film. We show that the He film thickness can be effectively
increased to become sufficient for the protection of neutrons from losses on
trap walls. In Sec. II we calculate the neutron wave function near a flat
trap wall covered by liquid He film as a function of neutron energy and of
film thickness. This calculation is straightforward and is required to correct the
quasiclassical formula (\ref{psiS1}) and to know what thickness of He film
is needed to protect UCN from any notable losses via trap walls. In Sec. III
we analyze the profile of He film on a flat vertical trap walls. In Sec. IV
we propose a method to increase the He film thickness. In Sec. V we
discuss the advantages, drawbacks, and possible prospects of using liquid
helium for improving material traps for the storage of UCN.

\section{Influence of helium film on the neutron absorption rate inside flat
trap walls}

In this section we calculate how the neutron wave function decreases inside
a flat trap wall covered by liquid He film. We need to determine the
minimal He film thickness required to protect the UCN from the absorption
inside the wall material. The reader, interested in only the qualitative
result of this paper, may skip the straightforward calculations of this
section and jump directly to its conclusions, given in Sec. IID. However,
for a quantitative study of UCN losses, the results obtained in this section
are important. In particular, that (i) the minimal He film thickness
required to protect the UCN is $\gtrsim d_{He}^{aim}\approx 100$nm, and (ii) Eq.
(\ref{psiS1}) $\approx 4$ times overestimates the reduction of UCN losses in a
wall covered by a helium film of thickness $d_{He}\gtrsim \kappa _{0He}^{-1}$
of our interest. 
\begin{figure}[tb]
\includegraphics[width=0.26\textwidth]{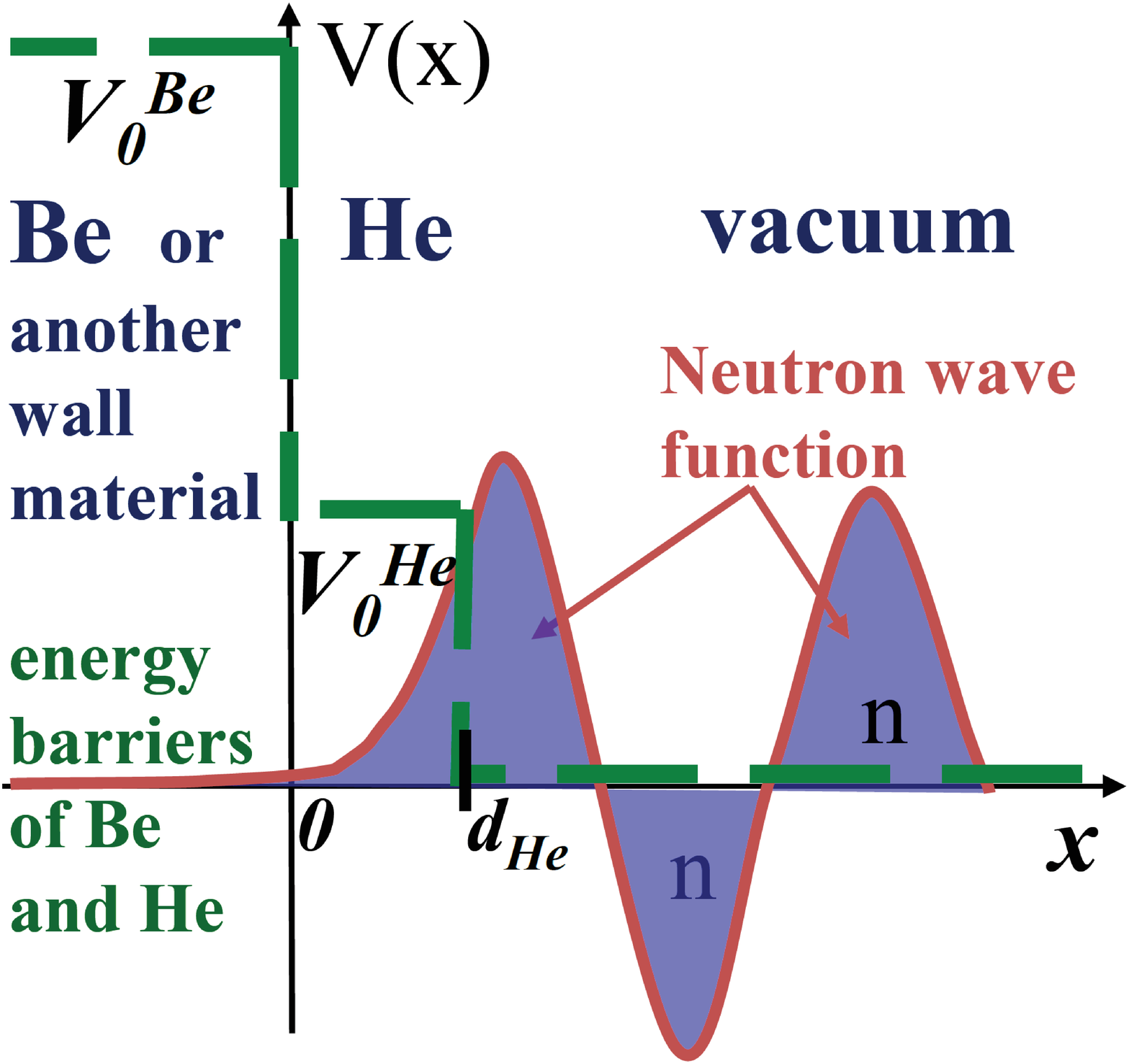}\ %
\includegraphics[width=0.21\textwidth]{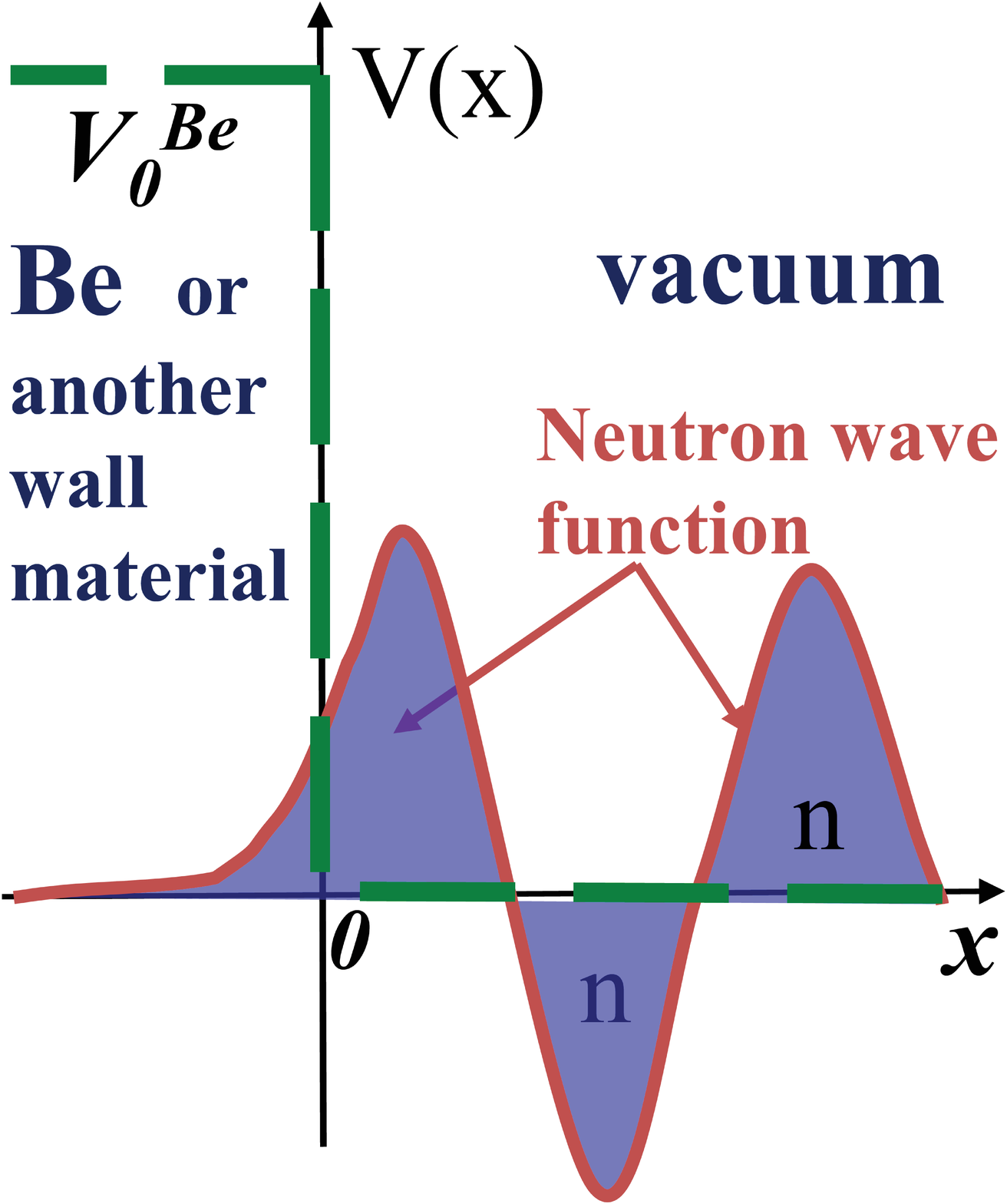}
\caption{Schematic representation of the ultracold neutron potential and
wave function near the trap wall, covered by He film (a) and without He film
(b). }
\label{FigScheme}
\end{figure}

\subsection{The model}

Consider the neutron wave function near the flat wall of a material neutron
trap. For an estimate of the influence of He film on neutron absorption rate
we study the one-dimensional quantum-mechanical problem, with only one
coordinate $x$. This 1D problem, of course, has been addressed before. In \S %
2.4.3 of the monograph \cite{Golub/1991} thin films absorbing neutrons were
studied. In contrast, here we consider the "insulating" non-absorbing He film and write
down explicit relations for the reduction factor of neutron absorption in
the material wall due to such a film. Depending on whether the trap wall is
covered by He film or not, we have the potential schematically shown in
Figs. \ref{FigScheme}a and \ref{FigScheme}b. In both cases inside the solid
wall at $x<0$, i.e. in the region I, the neutron wave function is given by%
\cite{LL3} 
\begin{equation}
\psi _{I}\left( x\right) =A\exp \left( \kappa _{W}x\right) ,~\kappa _{W}=%
\sqrt{2m_{n}\left( V_{0}^{W}-E\right) }/\hbar .  \label{psi1}
\end{equation}%
The neutron absorption rate $1/\tau _{a}$ inside the material wall for each
neutron state, given by its normalized wave function $\psi \left( x\right) $%
, is proportional to the probability $w_{W}$ of the neutron to be inside the
wall, i.e., at $x<0$: 
\begin{equation}
1/\tau _{a}\propto w_{W}=\int_{-\infty }^{0}dx\left\vert \psi \left(
x\right) \right\vert ^{2}=\left\vert A\right\vert ^{2}/\kappa _{W}.
\label{I}
\end{equation}%
Thus, to estimate the effect of He film on neutron storage time we need to
find the normalized neutron wave functions and their coefficients $A$ with
and without He film, and to compare the probabilities $w_{W}$ in Eq. (\ref{I}%
). We consider the case when the neutron energy $E=\hbar ^{2}k^{2}/2m$ is
smaller than the helium potential barrier $V_{0}^{He}$. Then inside helium
at $0<x<d_{He}$, i.e. in the region II, the neutron wave function 
\begin{eqnarray}
\psi _{II}\left( x\right) &=&B_{1}\exp \left( \kappa _{He}x\right)
-B_{2}\exp \left( -\kappa _{He}x\right)  \label{psi2m} \\
&=&B\sinh \left[ \kappa _{He}\left( x+x_{0}\right) \right] ,  \label{psi2}
\end{eqnarray}%
where 
\begin{equation}
\kappa _{He}=\sqrt{2m_{n}\left( V_{0}^{He}-E\right) }/\hbar =\sqrt{\kappa
_{0He}^{2}-k^{2}}.  \label{kHe}
\end{equation}%
Finally, in vacuum at $x>d_{He}$ the neutron wave function is\cite{LL3} 
\begin{equation}
\psi _{III}\left( x\right) =C_{1}\exp \left( ikx\right) +C_{2}\exp \left(
-ikx\right) =C\sin \left[ k\left( x+x_{1}\right) \right] ,  \label{psi3}
\end{equation}%
where $k\equiv \sqrt{2m_{n}E}/\hbar $.

\subsection{Wave function amplitudes}

\begin{figure}[tb]
\includegraphics[width=0.48\textwidth]{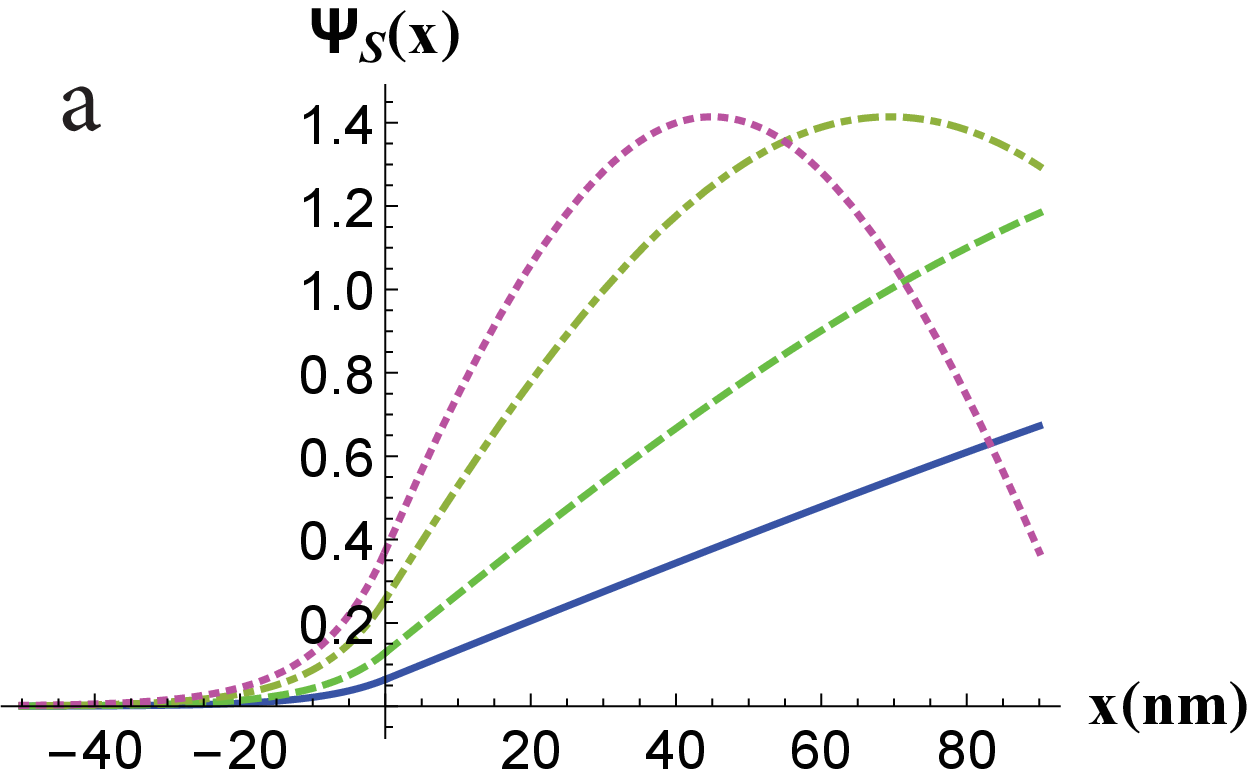}\newline
\medskip \includegraphics[width=0.48\textwidth]{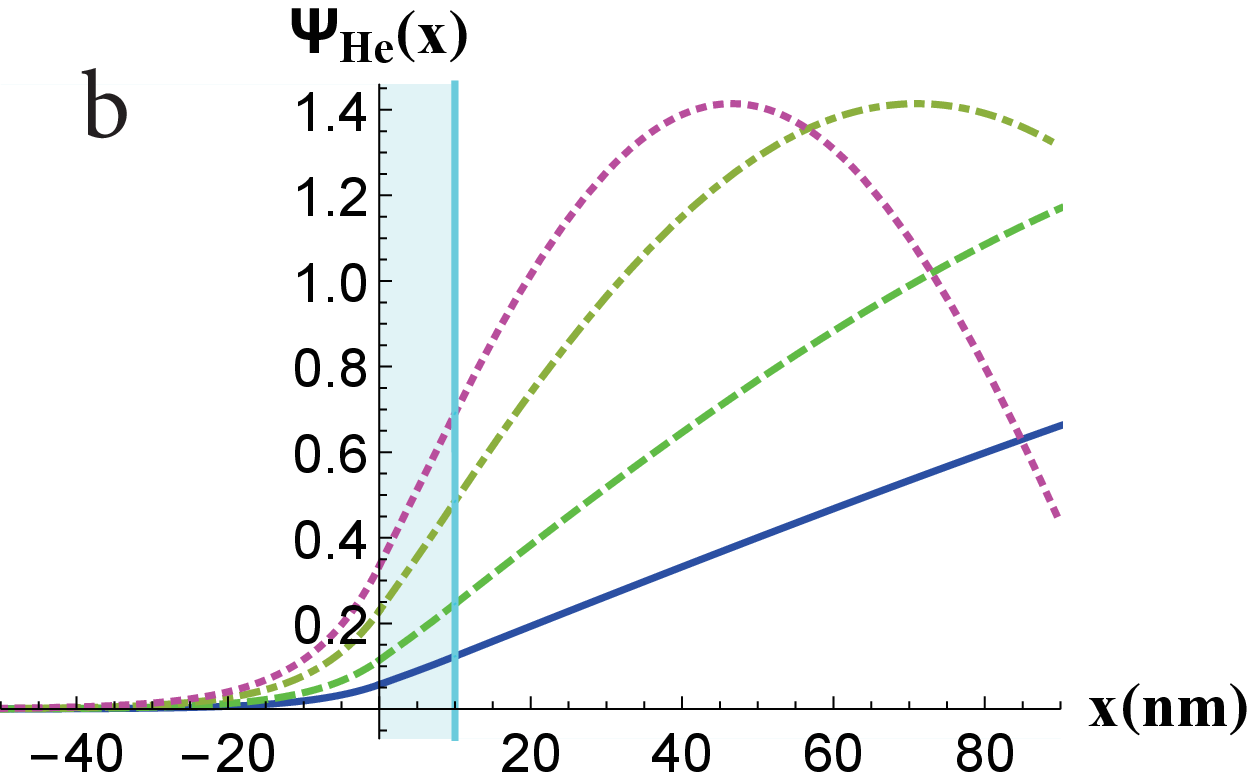}\newline
\medskip\includegraphics[width=0.48\textwidth]{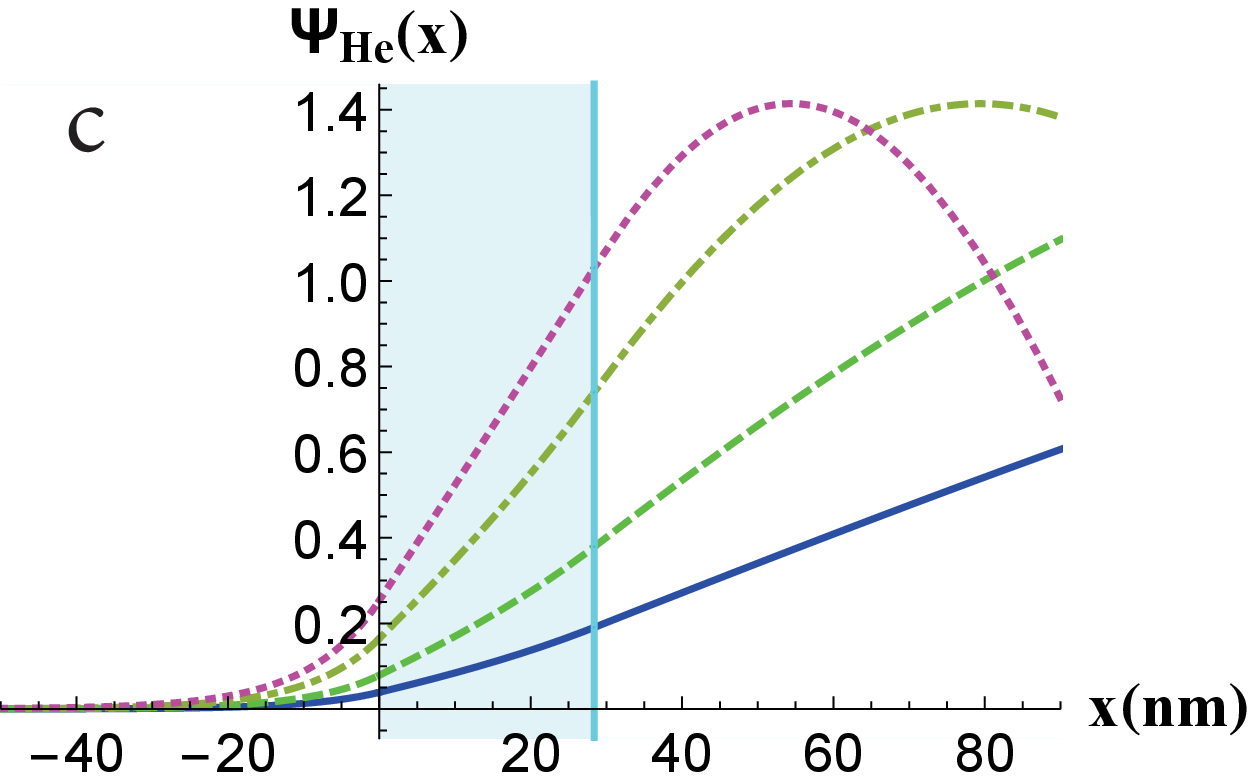}\ 
\caption{The neutron wave functions without (a) and with helium film of
thickness $d_{He}=10$nm (b) and $d_{He}=28.5$nm (c) at four values of
neutron wave vector: $k=0.005$ (solid blue), $0.01$ (dashed green), $0.02$
(dot-dashed orange), and $0.03$nm$^{-1}$ (dotted magenta).}
\label{FigWaveFunc}
\end{figure}

Without loss of generality, one can take $A$ to be real. Then, to satisfy
the standard boundary conditions (\ref{Boundaryx0}) and (\ref{BoundarydHe})
at $x=0$ and at $x=d_{He}$, requiring the continuity of the wave function
and its derivative, the coefficients $B_{1},B_{2},B,C,x_{0},x_{1}$ must also
be real. We assume the trap size $L$ $\gg d_{He},\kappa _{W}^{-1}$. Then the
normalization of neutron wave function for continuous spectrum, i.e. one
particle per unit volume, gives $C=\sqrt{2}$. Applying the boundary
conditions to the wave functions (\ref{psi1}),(\ref{psi2}) and (\ref{psi3}),
after the straightforward calculations given Appendix A, we obtain the
relations between wave-function amplitudes:%
\begin{eqnarray}
C &=&A\sqrt{\kappa _{W}^{2}/\kappa _{He}^{2}-1}\left\{ \left( \sinh \left[
\kappa _{He}\left( d_{He}+x_{0}\right) \right] \right) ^{2}\right. +  \notag
\\
&&+\left. \left( \cosh \left[ \kappa _{He}\left( d_{He}+x_{0}\right) \right]
\kappa _{He}/k\right) ^{2}\right\} ^{1/2},  \label{CA}
\end{eqnarray}%
\begin{equation}
B=A\sqrt{\kappa _{W}^{2}/\kappa _{He}^{2}-1},~  \label{BA}
\end{equation}%
and the coordinate shifts%
\begin{equation}
x_{0}=\frac{1}{2\kappa _{He}}\ln \left( \frac{\kappa _{W}+\kappa _{He}}{%
\kappa _{W}-\kappa _{He}}\right) \approx \frac{1}{\kappa _{W}},  \label{x0f}
\end{equation}%
\begin{equation}
x_{1}=\arctan \left( \tanh \left[ \kappa _{He}\left( d_{He}+x_{0}\right) %
\right] k/\kappa _{He}\right) /k-d_{He}.  \label{x1}
\end{equation}

For the "solid" trap wall without helium film, as shown in Fig. \ref%
{FigScheme}b, the region II is absent, and we only sew the wave functions (%
\ref{psi1}) and (\ref{psi3}) at $x=0$:%
\begin{equation}
A_{s}=C_{s}\sin \left[ kx_{1}\right] ;~A_{s}\kappa _{W}=C_{s}k\cos \left[
kx_{1}\right] ,  \label{As}
\end{equation}%
which gives%
\begin{eqnarray}
C_{s} &=&A_{s}\sqrt{1+\left( \kappa _{W}/k\right) ^{2}},~  \label{CAs} \\
x_{1s} &=&\frac{1}{k}\arcsin \left( \frac{1}{\sqrt{1+\left( \kappa
_{W}/k\right) ^{2}}}\right) .  \label{x1s}
\end{eqnarray}

The neutron wave functions at several $k$ without He film are shown in Fig. %
\ref{FigWaveFunc}a, for the minimal He film thickness $d_{He}^{\min }=10$nm
at the same $k$ in Fig. \ref{FigWaveFunc}b, and for He film thickness $%
d_{He}=28.5$nm, corresponding to the height $h=1$cm above He level, in Fig. %
\ref{FigWaveFunc}c. These figures illustrate the behavior of neutron wave
function $\psi \left( x\right) $ near the wall and show that $\psi
_{I}\left( x\right) $ inside the wall is strongly suppressed at small
neutron momentum $k$: $\psi \left( 0\right) $ $\propto 1/k$ at $k\ll \kappa
_{He}$. We also see that the thin He film due to Van der Waals attraction to
the walls does not change much the neutron wave function.

\subsection{Limiting cases}

At small neutron energy $E\ll V_{0}^{He}$, i.e., $k^{2}/\kappa _{He}^{2}\ll 1$%
, from Eq. (\ref{CA}) we get 
\begin{eqnarray}
C &\approx &B\kappa _{He}\cosh \left[ \kappa _{He}\left( d_{He}+x_{0}\right) %
\right] /k  \label{CBApp1} \\
&=&A\sqrt{\kappa _{W}^{2}-\kappa _{He}^{2}}\cosh \left[ \kappa _{He}\left(
d_{He}+x_{0}\right) \right] /k,  \label{CAApp1}
\end{eqnarray}%
and $x_{1}\approx x_{0}.$ If we also use $\kappa _{W}/\kappa _{He}\gg 1$,
substituting (\ref{x0f}) into (\ref{CAApp1}) we get 
\begin{eqnarray}
C &\approx &\frac{A\kappa _{W}}{k}\cosh \left[ \kappa _{He}d_{He}+\ln \sqrt{%
\frac{\kappa _{W}+\kappa _{He}}{\kappa _{W}-\kappa _{He}}}\right]  \notag \\
&\approx &\frac{A\kappa _{W}}{k}\cosh \left[ \kappa _{He}d_{He}+\kappa
_{He}/\kappa _{W}\right] .  \label{CAApp1t}
\end{eqnarray}%
At $E\ll V_{0}^{W}$, i.e. at $k/\kappa _{W}\ll 1$, Eqs. (\ref{CAs}) and (\ref%
{x1s}) also simplify: 
\begin{equation}
C_{s}\approx A_{s}\kappa _{W}/k;~x_{1s}\approx 1/\kappa _{W}.  \label{CAsApp}
\end{equation}%
At $\kappa _{He}d_{He}\ll 1$\ and $\kappa _{W}/\kappa _{He}\gg 1$, Eqs. (\ref%
{CAApp1t}) and (\ref{CAsApp}) coincide. From Eqs. (\ref{CAApp1}) and (\ref%
{CAApp1t}) it follows that the neutron absorption rate $w_{W}=\left\vert
A\right\vert ^{2}/\kappa _{W}$ in Eq. (\ref{I}), both with and without He
film, in the limit $k\rightarrow 0$ is strongly suppressed: $w_{W}\propto
k^{2}\propto E$. This agrees with the classical picture \cite%
{Ignatovich/1990} where the slow neutrons with speed $v$ have (i) longer
mean free time $\tau _{f}\sim L/v$ between hitting the walls, and (ii) spend
less time $\tau _{t}=m_{n}v/F$ interacting with the wall to reverse their
normal velocity due to the force $F$ acting from the wall.

In the opposite limit $E\rightarrow V_{0}^{He}$, from Eq. (\ref{kHe}) we
have $\kappa _{He}\rightarrow 0$, $k\rightarrow \kappa _{0He}$, and $%
k/\kappa _{He}\rightarrow \infty $. This gives 
\begin{equation}
x_{1}\approx \pi /2k-d_{He},
\end{equation}%
and from Eqs. (\ref{Eq1}) or (\ref{CA}) at $\kappa _{He}d_{He}\ll 1$ and $%
\kappa _{W}/\kappa _{He}\gg 1$ we get%
\begin{equation}
C\approx A\frac{\kappa _{W}}{k}\sqrt{1+k^{2}\left( d_{He}+x_{0}\right) ^{2}}.
\label{Copp}
\end{equation}%
At $kd_{He}\ll 1$ this coincides with Eq. (\ref{CAsApp}) without He.
However, at $\kappa _{He}^{-1}\gg d_{He}\gtrsim 1/k\approx \kappa
_{0He}^{-1} $ the He film reduces the neutron absorption rate $%
w_{W}=\left\vert A\right\vert ^{2}/\kappa _{W}$ several times, although
according to naive formula (\ref{psiS1}) its effect should be negligible.
This makes clear why at $k\rightarrow \kappa _{He}^{-1}\approx 30\mu m^{-1}$
the solid blue and dashed green curves in Fig. \ref{Figq}, corresponding to
Eq. (\ref{AA}), are always lower than the dot-dashed magenta curve
illustrating Eq. (\ref{psiS1}).

\subsection{The effect of He film covering the wall on neutron loss}

\begin{figure}[tbh]
{\Large {a}\includegraphics[width=0.47\textwidth]{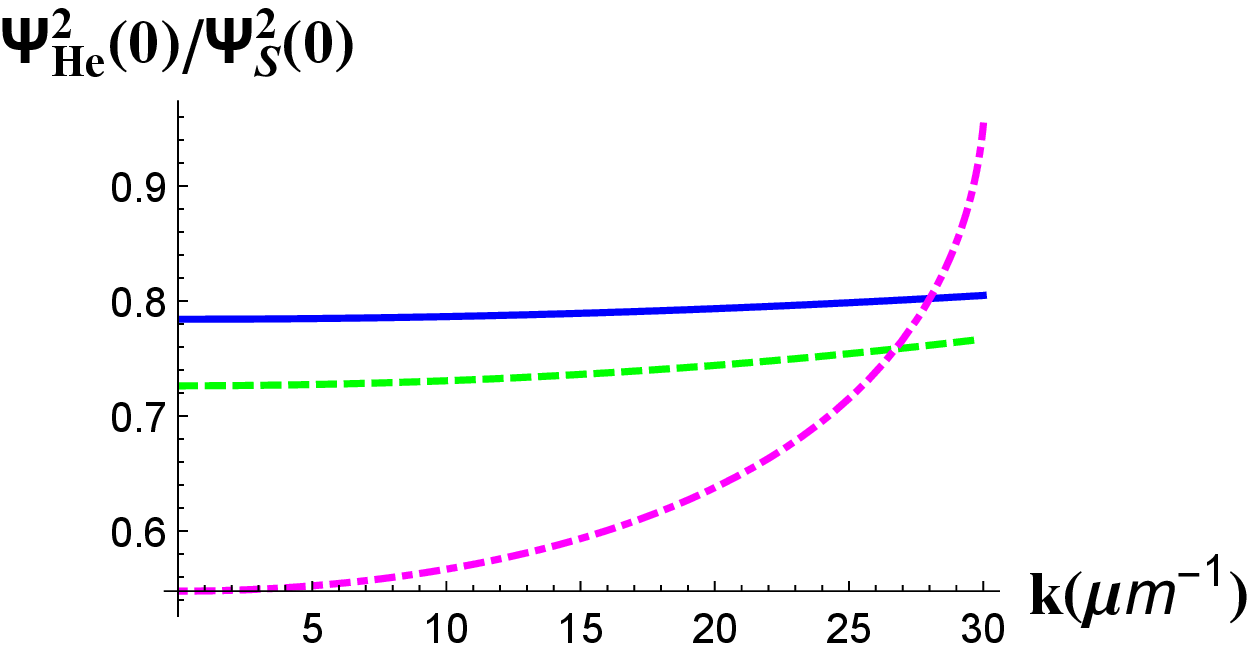} \bigskip\newline
{b}\includegraphics[width=0.47\textwidth]{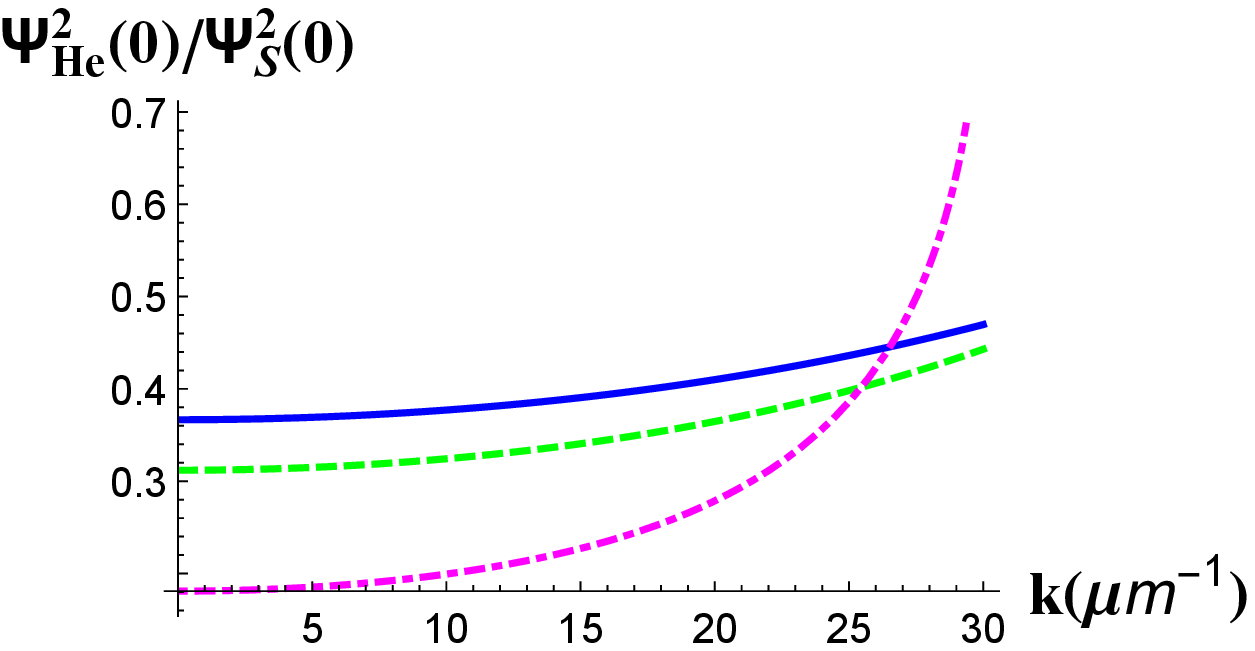}}
\caption{The reduction factor $\protect\gamma _{He}$ of neutron losses
inside flat trap walls, made of Be (solid blue) and \textquotedblright
Fomblin\textquotedblright\ oil (green dashed), due to the He film of
thickness $d_{He}=10$nm (a) and $d_{He}=28.5$nm (b). It illustrates the
momentum dependence of $\protect\gamma _{He}$ given by Eq. (\protect\ref{AA}%
). For comparison, the dot-dashed magenta curves show $\Psi
_{He}^{2}(0)/\Psi _{s}^{2}(0)$ determined using Eq. (\protect\ref{psiS1})
with $\protect\kappa _{He}$ given by Eq. (\protect\ref{kHe}).}
\label{Figq}
\end{figure}
Inside the absorbing wall the neutron wave function is given by Eq. (\ref%
{psi1}) in both cases, with and without He film covering the wall. However,
their amplitudes $A$ and $A_{s}$ differ, and so do the neutron loss rates $%
1/\tau _{a}$ given by Eq. (\ref{I}). Their ratio depends on the neutron
energy and on the He film thickness. Combining Eqs. (\ref{I}), (\ref{CA})
and (\ref{CAs}) at $C_{s}=C$, we obtain the reduction factor $\gamma _{He}$
of neutron absorption rates $1/\tau _{a}$ in the trap material due to the He
film:%
\begin{eqnarray}
\gamma _{He} &=&\frac{A^{2}}{A_{s}^{2}}=\frac{1+\left( \kappa _{W}/k\right)
^{2}}{\kappa _{W}^{2}/\kappa _{He}^{2}-1}\left\{ \left( \sinh \left[ \kappa
_{He}\left( d_{He}+x_{0}\right) \right] \right) ^{2}\right. +  \notag \\
&&+\left. \left( \cosh \left[ \kappa _{He}\left( d_{He}+x_{0}\right) \right]
\kappa _{He}/k\right) ^{2}\right\} ^{-1},  \label{AA}
\end{eqnarray}%
where $x_{0}$ is given by Eq. (\ref{x0f}). The ratio $\gamma _{He}\equiv
A^{2}/A_{s}^{2}=\Psi _{He}^{2}(0)/\Psi _{s}^{2}(0)$ as a function of $k$ for 
$\kappa _{0W}=\kappa _{0Be}$ and $\kappa _{0W}=\kappa _{0F}$ is shown in
Fig. \ref{Figq}. In Fig. \ref{Figq}a we used $d_{He}=10$nm, corresponding to
the minimal He film thickness. One sees that this ratio is close to unity,
which means that the effect from a very thin He film due to the additional
suppression of the neutron wave function is not very important for neutron loss
rate. However, in Fig. \ref{Figq}b we show $A^{2}/A_{s}^{2}$ for the same
set of $k$ but at $d_{He}=28.4$nm, corresponding to the He film thickness at
height $h=1cm$. The effect of He film at $d_{He}=28.4$nm is already
considerable, as it reduces the neutron absorption rate $\approx 3$times.

\begin{figure}[tbh]
{\Large \includegraphics[width=0.48\textwidth]{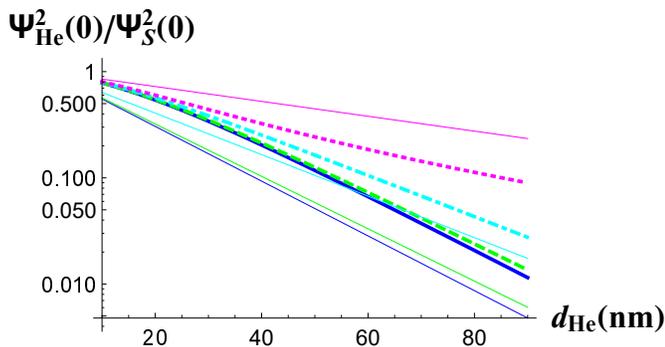}\bigskip \newline
}
\caption{The squared ratio (in logarithmic scale) of neutron wave functions
on the wall surface, made of Be, as a function of helium film thickness $%
d_{He}$ at four different values of neutron wave vector $k=0.005$ (solid
blue), $0.01$ (dashed green), $0.02$ (dot-dashed cyan), and $0.03$nm$^{-1}$
(dotted magenta). Thin solid curves of the corresponding colors show the
predictions of Eq. (\protect\ref{psiS1}).}
\label{FigqL}
\end{figure}

In the limit $k^{2}\ll \kappa _{He}^{2}$ Eq. (\ref{AA}) simplifies to 
\begin{equation}
\frac{A^{2}}{A_{s}^{2}}=\cosh ^{-2}\left[ \kappa _{He}d_{He}+\frac{1}{2}\ln 
\sqrt{\frac{\kappa _{W}+\kappa _{He}}{\kappa _{W}-\kappa _{He}}}\right]
/\left( 1-\frac{\kappa _{He}^{2}}{\kappa _{W}^{2}}\right) .  \label{AAs}
\end{equation}%
This absorption-rate ratio is independent of $k$, because both $A^{2}$ and $%
A_{s}^{2}\propto k^{2}$. Comparing Eqs. (\ref{AAs}) and (\ref{psiS1}) we
obtain that the latter gives the absorption rate $1/\tau _{a}$ smaller than
the correct result by a factor of $4$ at $\kappa _{He}d_{He}\gg 1$ and $%
k\ll \kappa _{He}\ll \kappa _{W}$, which is also seen from Fig. \ref{FigqL}.

The dependence of $\gamma _{He}\equiv A^{2}/A_{s}^{2}$ on the film thickness 
$d_{He}$ at several $k$ is shown in Fig. \ref{FigqL} for a beryllium wall. For
the trap walls covered by \textquotedblleft Fomblin\textquotedblright\ oil
the dependence $A^{2}/A_{s}^{2}$ on $d_{He}$ looks very similar to Fig. \ref%
{FigqL} and, hence, is not shown here. For convenience the plot is given in
logarithmic scale. One sees that the expected exponential suppression (\ref%
{psiS1}) of neutron probability density inside liquid helium starts from
film thickness $\approx 20$nm. For comparison, by thin solid lines of the same
colors in Fig. \ref{FigqL} we also show the predictions of Eq. (\ref{psiS1}%
), which are quantitatively incorrect. At small $k$ Eq. (\ref{psiS1}) gives
a neutron absorption rate $\approx 4$ times smaller than the exact result (%
\ref{AA}). At large $k\rightarrow \kappa _{0He}$, in contrast, Eq. (\ref%
{psiS1}) predicts a too large absorption rate, as follows from Eq. (\ref{Copp}%
) and discussed after it.

The simple calculations of this section show several points, important for
possible application of He films to reduce neutron losses inside a material
trap. First, the approximate estimate in Eq. (\ref{psiS1}) works
qualitatively but not quantitatively. At low neutron energy $E<0.7V_{0}^{He}$%
, which is of main interest, the calculated $\gamma _{He}$ in Eq. (\ref{AA})
is up to four times smaller than $\gamma _{He}$ predicted from Eq. (\ref%
{psiS1}), as follows from Eq. (\ref{AAs}) and is shown in Figs. \ref{Figq}
and \ref{FigqL}. As illustrated in Fig. \ref{FigqL}, this discrepancy
increases with film thickness and reaches four times at $d_{He}\gg \kappa
_{He}^{-1}$. At large energy $E\rightarrow V_{0}^{He}$, on contrary, Eq. (%
\ref{psiS1}) predicts too small reduction factor $\gamma _{He}$, as follows
from Eq. (\ref{Copp}). Especially, this is clear at $E\geq V_{0}^{He}$,
where according to Eq. (\ref{psiS1}) $\gamma _{He}=1$, because the imaginary
part of neutron momentum is zero. This discrepancy at $E\gtrsim V_{0}^{He}$
appears because Eq. (\ref{psiS1}) neglects the backscattering of neutron
waves from the He surface, which takes place even if $E>V_{0}^{He}$.

The second, physical conclusion from the above calculations is that a thin
He film of thickness $10-30$nm, formed only by the Van der Waals attraction
of superfluid helium to the trap walls, is not sufficient to completely
avoid the neutron losses due to the absorption inside the wall material. The
neutron absorption rate $1/\tau _{a}$ inside an ideal flat wall reduces due
to such a thin He film only by a factor $\gamma _{He}\equiv $ $%
A^{2}/A_{s}^{2}\sim 0.3-0.8$ (see Figs. \ref{Figq} and \ref{FigqL}),
depending on the film thickness determined by the height $h$ above He level
(see next section). As one sees from Fig. \ref{FigqL}, to reduce a hundred
times the neutron losses due to the absorption in material walls by
covering with liquid helium, the He film thickness must be $d_{He}\gtrsim
d_{He}^{aim}=100$nm.

\section{Helium film profile on a flat vertical wall}

The study of the profile of superfluid helium film on vertical walls,
i.e. the dependence of He film thickness $d_{He}$\ on the height $z$\ above
liquid helium level, turned out to be a non-trivial problem, both
experimentally\cite%
{Atkins1950,BurgeJackson1951,RaymondBowers,Atkins1957,PhysRevA.7.790,PhysRevA.9.1312}
and theoretically\cite%
{Lifshitz1955,Atkins1957,PhysRevA.7.790,PhysRevA.9.1312}.

According to classical fluid mechanics,\cite{LL6} the meniscus profile
is given by 
\begin{eqnarray}
\frac{d_{He}\left( h\right) }{a_{He}} &=&\arccosh\left( \frac{2a_{He}}{h}%
\right) -\arccosh\left( \frac{2a_{He}}{h_{0}}\right)  \notag \\
&&-\sqrt{4-h^{2}/a_{He}^{2}}+\sqrt{4-h_{0}^{2}/a_{He}^{2}},  \label{Meniscus}
\end{eqnarray}%
where the capillary length of liquid $^{4}$He is $a_{He}=\sqrt{\sigma
_{He}/g\rho _{He}}=0.5$mm and $h_{0}=\sqrt{2}a_{He}\sqrt{1-\sin \theta }$ is
the maximal height to which the fluid rises at the wall. Here $\sigma
_{He}=0.354$ dyn/cm is the surface tension coefficient of liquid $^{4}$He, $%
g=9.8$m/s$^{2}$, and $\rho _{He}\approx 0.145$ g/cm$^{3}$. Even for the zero
contact angle $\theta $, according to Eq. (\ref{Meniscus}), the film
thickness $d_{He}=0$ for $h>h_{0}$. This is not the case for superfluid
helium that covers the walls and ceiling of the trap by a thin film of
thickness $d_{He}^{\min }\approx 10$nm at arbitrary height $h$ because of
the Van der Waals attraction to the walls. However, the correct profile of
helium film is not just the sum of $d_{He}\left( h\right) $ in Eq. (\ref%
{Meniscus}) and the minimal film thickness $d_{He}^{\min }\approx 10$nm.

The simple theory of the superfluid He film profile at $h>h_{0}$ assumes
that the total energy $E_{tot}$ of a $^{4}$He atom on the surface of He film
in equilibrium does not depend on the height $h$. This total energy contains
the gravitational energy $M_{He}gh$ of this atom and its Van der Waals
attraction $V_{W}$ to the wall. As a result, one obtains an equation
relating $d_{He}$ and $h$: 
\begin{equation}
E_{tot}\left( h\right) =M_{He}gh+V_{W}(d_{He})=0.  \label{Etot}
\end{equation}%
The Van der Waals attraction to other He atoms is assumed to be the same on
the surface of thin He film on the wall and on the thick film covering the
vessel bottom. Therefore, it is omitted in Eq. (\ref{Etot}). The microscopic
theory giving 
\begin{equation}
V_{W}(d_{He})\propto d^{-n}  \label{Lif55}
\end{equation}%
was developed in Ref. \cite{Lifshitz1955}, where $n=3$ or $4$ depending on
the distance and wall material. Combining Eqs. (\ref{Etot}) and (\ref{Lif55}%
) gives%
\begin{equation}
d_{He}\left( h\right) =d_{0}/h^{1/n}.  \label{dHe}
\end{equation}

\begin{figure}[tbh]
\includegraphics[width=0.48\textwidth]{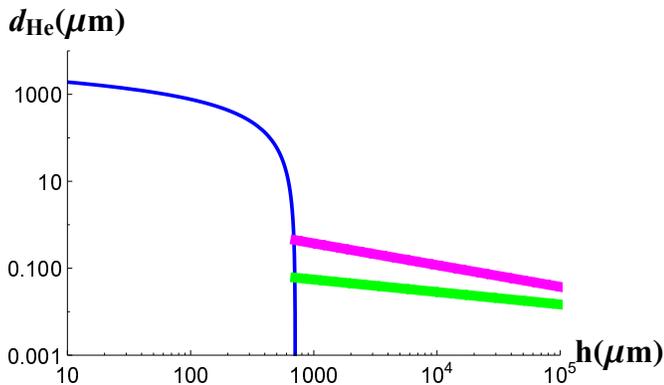}
\caption{The functions $d_{He}\left( h\right) $ from Eq. (\protect\ref%
{Meniscus}) (solid blue curve) and from Eq. (\protect\ref{dHe}) with the
parameters\protect\cite{PhysRevA.9.1312} $d_{0}=28.5nm$ and $1/n\approx
1/3.5 $ (dashed green line) and\protect\cite{RaymondBowers} $d_{0}=118nm$
and $1/n\approx 0.5$ (dash-dotted magenta line).}
\label{FigdHe}
\end{figure}

The experiments\cite%
{Atkins1950,BurgeJackson1951,RaymondBowers,Atkins1957,PhysRevA.7.790,PhysRevA.9.1312}
confirmed Eq. (\ref{dHe}), but the values of $d_{0}$ and of the exponent $%
1/n $ slightly vary depending on the measurement method. If one denotes $%
d_{0}\equiv d_{He}\left( h=1cm\right) $, and the height $h$ is also
given in units of cm, the latest measurement\cite{PhysRevA.9.1312} using the He
oscillation method suggests $d_{0}=28.5nm$ and $1/n\approx 1/3.5=0.286$.
This reasonably agrees with earlier measurements using He oscillations\cite%
{Atkins1950} and optical methods\cite{BurgeJackson1951,Ham1954}. The latter
suggests\cite{Ham1954} slightly different values $d_{0}=30$nm and $n\approx
2-3$. At smaller $h$ the exponent increases\cite{RaymondBowers,Atkins1957}
to $1/n\approx 0.5$. For example, measurements using the microbalance weight
method at $h>3mm$ at temperature $1.2K<T<2.13K$ suggest\cite{RaymondBowers}
much thicker film and the parameters $d_{0}\approx 118nm$ and $n\approx 2.$
The functions $d_{He}\left( h\right) $ from Eqs. (\ref{Meniscus}) and (\ref%
{dHe}) are plotted in Fig. in logarithmic scale. As one sees from this plot,
the sewing of Eqs. (\ref{Meniscus}) and (\ref{dHe}) is still lacking,
because, unfortunately, we could not find data on $d_{He}\left( h\right) $
in the region $h_{0}<h<3mm$, while Eq. (\ref{Meniscus}) is not valid there.
Probably, in this region the exponent in Eq. (\ref{dHe}) further increases
to $1/n>0.5$. Figure \ref{FigdHe} shows the general dependence $d_{He}\left(
h\right) $ and can be used for the estimates of neutron absorption rate on a
flat vertical wall covered by superfluid helium.

Although the He meniscus is wide enough to completely protect neutrons from
absorption inside a solid wall material, according to Eq. (\ref{AA}), its
height is too small. Neutrons with energy $E=E_{\max }=V_{0}^{He}=18.5$%
neV in the eart'sh gravity potential may reach the height $h_{\max
}=E/m_{n}g\approx 18$cm. The meniscus height is only $h_{m}\approx 0.7$mm,
i.e., much smaller. Hence, most neutrons with energy $E<V_{0}^{He}$ heat
the trap walls much higher than He meniscus, where the thickness of He film
covering an ideal flat vertical wall is only determined by the Van der Waals
forces.

\section{He film on a rough wall}

The wall roughness, usually, enhances $2$-$3$ times the neutron losses due
to the absorption inside trap walls.\cite{Ignatovich/1990,Golub/1991} This
happens because the wall roughness makes the repulsion potential of the
walls smoother, so that the neutron wave function penetrates deeper into the
walls.\cite{Ignatovich/1990,Golub/1991} However, the wall roughness may
considerably increase the average He film thickness due to capillary
effects, thus reducing the neutron losses. Indeed, the wall roughness
increases its surface area, raising the role of capillary effects. If the
length scale of surface roughness $l_{R}\ll a_{He}$, to minimize the surface
tension energy the He film even on a rough wall must have almost flat
interface with vacuum. Hence, the superfluid helium fills all small cavities
of size $l_{R}\lesssim a_{He}$ in the wall.

To describe the He profile one has to minimize the energy functional of He
film, 
\begin{equation}
E_{tot}=V_{g}+E_{s}+V_{W},  \label{EtotM}
\end{equation}%
instead of considering a single He atom, as we did in Eq. (\ref{Etot}). Here
the gravity term is 
\begin{equation}
V_{g}=\rho _{He}g\int zd_{He}\left( \boldsymbol{r}_{||}\right) \,d^{2}%
\boldsymbol{r}_{||},  \label{Vg}
\end{equation}%
where $\boldsymbol{r}_{||}=\left\{ y,z\right\} $ is a 2D coordinate vector
along the wall plane, $y$ and $z$ are the horizontal and vertical
coordinates along the wall, 
\begin{equation}
d_{He}\left( \boldsymbol{r}_{||}\right) =\xi \left( \boldsymbol{r}%
_{||}\right) -\xi _{W}\left( \boldsymbol{r}_{||}\right)   \label{dHeDef}
\end{equation}%
is the coordinate dependent He film thickness, and the functions $\xi \left( 
\boldsymbol{r}_{||}\right) $ and $\xi _{W}\left( \boldsymbol{r}_{||}\right) $
describe the surface profiles of He and of trap wall. Below we consider a
wall roughness with typical length scale $\lesssim a_{He}\ll h_{\max }$. The
variation of the coordinate $z$ on this small length scale can be neglected
compared to its average $\left\langle z\right\rangle $, i.e., its height $%
h=\left\langle z\right\rangle $ above the He level. Hence, in Eq. (\ref{Vg})
the coordinate $z$ can be replaced by the height $h$ of the wall roughness. 

The second term in Eq. (\ref{EtotM}), describing the surface tension energy,
is given by 
\begin{equation}
E_{s}=\sigma _{He}\int \sqrt{1+\left[ \boldsymbol{\nabla }\xi \left( 
\boldsymbol{r}_{||}\right) \right] ^{2}}d^{2}\boldsymbol{r}_{||}.  \label{Es}
\end{equation}%
Its square-root dependence complicates the problem of finding an exact
surface profile $\xi \left( \boldsymbol{r}_{||}\right) $. Usually, its
analytical solution is available only in the limit $\left\vert \boldsymbol{%
\nabla }\xi \left( \boldsymbol{r}_{||}\right) \right\vert \ll 1$.

The gravity and surface tension, i.e., the first two terms in Eq. (\ref{Etot}%
), are important on a macroscopic length scale $\gtrsim a_{He}$. The last
van der Waals term $V_{W}$, describing the helium attraction to the wall
material, acts on a much shorter distance $\lesssim d_{He}^{\min }\approx
10nm\ll a_{He}$. We have a very lucky situation for a theoretical analysis,
because the van der Waals length scale $\sim d_{He}^{\min }$ is five orders
of magnitude smaller than the capillary length scale $a_{He}$. Hence, the
influence of gravity and of the surface tension of free He surface on $V_{W}$
can be neglected. For a flat wall surface the van der Waals term $V_{W}$
depends only on the wall material and on the film thickness: $%
V_{W}=V_{W}(d_{He})$. Without surface tension it would lead to the covering
of a rough surface by a He film of thickness $d_{He}\sim $ $d_{He}^{\min }$,
which almost repeats the wall profile if $l_{R}\gg d_{He}^{\min }$. Thus one
may keep only two first terms in the functional $E_{tot}\left[ \xi \left( 
\boldsymbol{r}_{||}\right) \right] $ in Eq. (\ref{EtotM}), reducing the
effect of van-der-Waals term $V_{W}$ to the "boundary conditions" of the
minimal He film of thickness $d_{He}\sim $ $d_{He}^{\min }$.

Such a minimal He film of thickness $d_{He}\sim $ $d_{He}^{\min }$, caused by
the van der Waals attraction, would cost an additional surface tension
energy $\Delta E_{s}$, which may be larger than the gravity term $\Delta
V_{g}$ of extra helium needed to make the He surface even flat: $\xi \left( 
\boldsymbol{r}_{||}\right) =const=\max \left\{ \xi _{W}\left( \boldsymbol{r}%
_{||}\right) \right\} +d_{He}^{\min }$. This additional amount of helium
depends on the wall profile and may strongly increase the average thickness
of He films. Below we consider several types of surface roughness. We do
not calculate exact surface profiles $\xi \left( \boldsymbol{r}%
_{||}\right) $ for particular functions $\xi _{W}\left( \boldsymbol{r}%
_{||}\right) $, but make some simple qualitative estimates of the
roughness parameters when the surface tension leads to almost flat He
surface $\xi \left( \boldsymbol{r}_{||}\right) $.

\subsection{Hemispherical cavity}

Consider a semispherical cavity of radius $R$, $d_{He}^{\min }\ll R\lesssim
a_{He}$, located in a wall at a height $h$ above the liquid He level. Due to
the van der Waals forces, its surface is covered by superfluid helium film.
However, due to capillary effects, for small $R\lesssim a_{He}^{2}/h$ this
cavity is almost totally filled with helium. Indeed, a thin He film of
thickness $d_{He}$ on the surface of a hemisphere has the total surface $%
S=2\pi \left( R-d_{He}\right) ^{2}\approx 2\pi R^{2}$, and the corresponding
surface tension energy loss is $E_{s1}\approx 2\pi R^{2}\sigma _{He}$. If
this cavity is totally filled with He, this surface tension energy reduces
to $E_{s2}\approx \pi R^{2}\sigma _{He}$, but the gravitational energy
increases by $E_{g}\approx 2\pi R^{3}\rho _{He}gh/3$. Hence, it is
energetically favorable to fill the cavity by helium if $E_{g}<\Delta
E_{s}\equiv E_{s1}-E_{s2}$, or if 
\begin{equation}
R<R_{0}\left( h\right) \approx 3\sigma _{He}/2\rho _{He}gh=3a_{He}^{2}/2h.
\end{equation}%
For $h=h_{\max }=18$cm, this gives $R_{0}\approx 2\mu m$. Helium film of
such thickness is more than enough to protect neutrons from absorbing inside
the material wall. Of course, the exact He surface profile inside a
hemispherical cavity is not spherical or flat, and even for $R>R_{0}\left(
h\right) $ the cavity will be partially filled with liquid helium, so that
the film thickness $d_{He}\gg \kappa _{He}^{-1}$. An additional and important
advantage of He film on the walls is that it fills the surface cavities,
thus reducing the neutron losses caused by surface roughness.

The above simple estimate demonstrates the possibility of a strong increase
of the effective thickness of He film on the trap walls. The natural surface
roughness depends on materials and their preparation, which is a subject of
much investigation \cite{whitehouse2004surfaces}. We propose to create the
special roughness of trap walls to increase the effective thickness of He
film and, hence, to reduce the neutron losses.

\subsection{Ribbed wall surface}

\begin{figure}[tbh]
\includegraphics[width=0.48\textwidth]{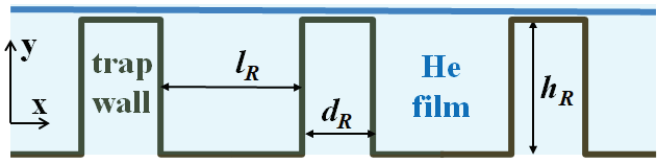}
\caption{The steplike roughness of the wall surface covered by liquid
helium with an almost flat surface.}
\label{FigStep}
\end{figure}

Now consider a periodic 1D roughness of an asymmetric square-wave form, i.e.,
of straight rectangular protrusions (ridges) as shown in Fig. \ref{FigStep}.
Again we compare the energies of two He configurations: (i) thin He film
covering the wall surface due to van der Waals term $V_{W}$ and (ii) flat
surface completely covering the surface roughness. The estimates are similar
to those in the previous subsection. The gravity energy per unit length is $%
E_{g}\approx \rho _{He}ghh_{R}l_{R}$, where $h$ is again the height on the
wall, $h_{R}$ is the roughness height, and $l_{R}$ is the distance between
the ridges. The surface energy difference for these two configurations per
unit length is $\Delta E_{s}\approx h_{R}\sigma _{He}$. It is energetically
favorable to fill the roughness by helium if $\Delta E_{s}>E_{g}$, or if
\begin{equation}
l_{R}<l_{0}\left( h\right) \approx \sigma _{He}/\rho _{He}gh=a_{He}^{2}/h.
\label{lR}
\end{equation}%
The roughness height $h_{R}$ can be made larger than $d_{He}^{aim}=100$nm.
Then the only neutron absorption in the wall material is due to the
rectangular protrusions. However, the volume part $\phi $ of such
protrusions is small, $\phi \approx d_{R}/l_{R}\ll 1$, where $d_{R}$ is the
ridge width (see Fig. \ref{FigStep}). Using Eq. (\ref{lR}) we obtain the
minimal value of this volume fraction, $\phi _{\min }\approx
d_{R}h/a_{He}^{2} $. If one takes the width of ridges $d_{R}\approx
0.1d_{He}^{aim}=10$nm, which is sufficient for the full van der Waals
attraction of He to the wall, and $h=h_{\max }=18$cm, one obtains 
$\phi _{\min }\approx d_{R}h_{\max }/a_{He}^{2}=7.2\times 10^{-3}$, 
i.e., more than a hundred times
reduction of neutron losses due to the absorption inside wall material.

Making the height of protrusions on the wall ten times higher that
their width looks technically difficult because of their fragility. To raise
the durability of surface roughness, one may use zig-zag instead of
straight grooves. Taking thicker ridges with $d_{R}\approx d_{He}^{aim}=100$%
nm gives $\phi _{\min }\approx 0.07$ and the neutron loss reduction factor $%
\gamma \approx 1/14$. This is also good. To further increase the efficiency
of He film protection one may consider other roughness configurations. One
possibility is to make height-dependent linear density of ridges $%
1/l_{R}\propto h$. According to Eq. (\ref{lR}), $1/l_{R}>1/R_{0}\left(
h\right) =h/a_{He}^{2}$. Making $1/l_{R}=h/a_{He}^{2}$ instead of $%
1/l_{R}=h_{\max }/a_{He}^{2}$ reduces the protrusion density twice, thus
diminishing the neutron losses due to the wall absorption by a factor $\approx
28$.

Note that the required small period $l_{R}= a_{He}^{2}/h_{\max}\approx 1.4\mu m$ 
of surface roughness is technically achievable by many methods. Diffraction 
gratings of this and even smaller periods are commercially available. 
Even a much shorter period $l_{R} \leq 100$nm of the grating is technically 
possible with a rather high precision by using electron-beam lithography.\cite{Bourgin2010} 
The coverage of UCN trap walls by the powder of diamond nanoparticles, 
often used in experiments with UCN \cite{DiamondPowder}, may also create 
the required surface roughness to make the helium film sufficiently thick. 

\subsection{Fur surface roughness}

\begin{figure}[tbh]
\includegraphics[width=0.48\textwidth]{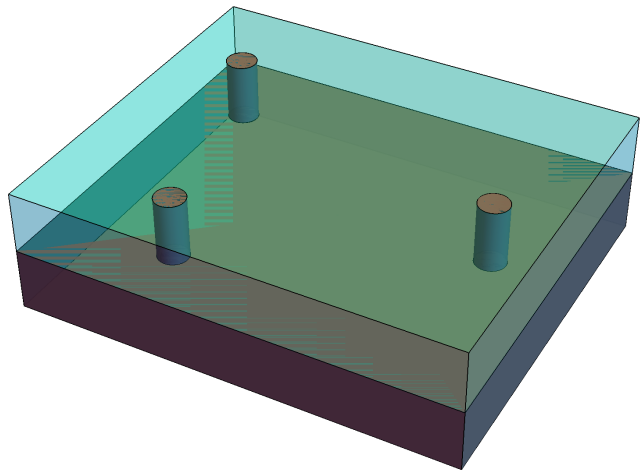}
\caption{Schematic view of a furry rough wall surface covered by liquid
helium.}
\label{FigFur}
\end{figure}
Another way to optimize the surface roughness is to make it
\textquotedblright furry\textquotedblright . This type of disorder on
neutron trap walls was much investigated theoretically \cite%
{Ignatovich/1990,Golub/1991}, because it corresponds to a random surface
roughness with short-range Gaussian correlation function. We approximate
this roughness by randomly situated thin cylinders of height $h_{R}$,
diameter $d_{R}\gg d_{He}^{\min }$, and surface density $n_{R}$, protruding
from the flat trap wall, as shown in Fig. \ref{FigFur}. Each cylinder
increases the surface area by $\Delta S\approx \pi d_{R}h_{R}$, and covering
the fur with He film reduces the surface tension energy by $\Delta
E_{s}\approx \pi d_{R}h_{R}\sigma _{He}n_{R}$. The gravitational energy of
such a He film per unit wall area is $E_{g}\approx \rho _{He}ghh_{R}$. It is
energetically favorable to fill the \textquotedblright
fur\textquotedblright\ roughness by liquid helium if $\Delta E_{s}>E_{g}$,
or if the fur density%
\begin{equation}
n_{R}>n_{R}\left( h\right) =h/\pi d_{R}a_{He}^{2}.  \label{nR}
\end{equation}%
Taking this fur density $n_{R}\left( h\right) =h/\pi d_{R}a_{He}^{2}$, one
obtains the volume fraction $\phi $ of such protrusions in the He film to be 
\begin{equation}
\phi =\int_{0}^{h_{\max }}\left( \pi d_{R}^{2}/4\right) n_{R}\left( h\right)
dh/h_{\max }=d_{R}h_{\max }/8a_{He}^{2}.
\end{equation}%
Taking the cylinder diameter equal to its height, $d_{R}=h_{R}\approx
d_{He}^{aim}=0.1\mu m$, one obtains that the neutron losses due to the wall
absorption can be reduced by a factor $\gamma =1/\phi \approx 110$. Even a
much lower surface roughness, when the cylinder height is equal to only $%
1/10 $ of its diameter, reduces the neutron losses by a factor $>10$.
Taking into account another benefit from using He film, namely, the
elimination of negative effect of surface roughness on neutron losses, this
shows a strong advantage of using superfluid He to cover the walls of a
neutron trap.

Before ending this section we note that the surface roughness and the
corresponding increase of the effective He film thickness due to capillary
effects may explain the discrepancy of experimental values determined by
different methods \cite%
{Atkins1950,BurgeJackson1951,RaymondBowers,Atkins1957,PhysRevA.7.790,PhysRevA.9.1312}%
. Indeed, the microbalance weight method \cite{RaymondBowers} measures the
total He weight. This He weight includes the filled cavities on the surface,
which are not detected by optical methods \cite{BurgeJackson1951,Ham1954}.
Hence, the weight method gives a thicker He film on a rough surface\cite%
{RaymondBowers}. Since the typical He film thickness due to van der Waals
forces is small, $d_{He}\sim 10$nm, even a tiny surface roughness of height $%
\sim 10$nm may strongly affect the measured values of He film.

\section{Discussion}

$^{4}$He does not absorb neutrons and can be used to protect neutrons from
the absorption inside the material of trap walls, thus considerably increasing 
the neutron storage time. However, liquid He
introduces new scatterer for neutrons: the He vapor atoms and the soft
thermal excitations - the quanta of surface waves, called ripplons. Let us
summarize the benefits and disadvantages of covering the UCN trap walls with
liquid helium.

\emph{I. Drawbacks.}

1. Liquid $^{4}$He has a very small optical potential barrier $%
V_{0}^{He}=18.5\ \mathrm{neV}$ for the neutrons, which is about one order of
magnitude smaller than most of other materials used for UCN solid traps.
Hence, only UCN with kinetic energy $E<V_{0}^{He}$ can be effectively stored
in such a trap. Hence, the corresponding UCN density inside a He trap is about
one order of magnitude smaller than inside traps made of other materials.
Unfortunately, the only way to overcome this drawback is to increase the
neutron density coming from their source. The simple neutron manipulations
using any external potential cannot increase the density of neutrons with
energy $E<V_{0}^{He}$ because of the Liouville's theorem, stating the
conservation of particle density in phase space for Hamiltonian systems.
However, modern neutron reactors give rather high neutron intensity.\cite%
{PhysRevC.95.045503} Hence, the reduction of UCN density in a trap by an
order of magnitude is important but not crucial. More complicated neutron
manipulations, e.g., using inelastic magnetic scattering in weakly absorbing
cold paramagnetic systems \cite{ZimmerPhysRevC.93.035503}, probably, may
further increase the density of UCN.

2. A thick He film covers only the bottom of the neutron trap. Due to
capillary effects, the flat vertical walls on the height $h<a_{H}\sqrt{2}%
\approx 0.7mm$ is also covered by liquid helium of sufficient thickness $%
d_{He}\sim a_{H}$. However, the maximal height $h_{\max }$ to which the UCN
with kinetic energy $E<V_{0}^{He}$ may fly is much larger than $a_{H}\sqrt{2}
$: $h_{\max }\approx 18$cm. Above the height $a_{H}\sqrt{2}$ the superfluid
He covers the flat walls due to the van der Waals attraction, but the
corresponding He film thickness $d_{He}=10-30$nm is not sufficient to
protect the neutrons from their penetration into the wall material. One
needs a thicker He film $d_{He}>d_{He}^{aim}\approx 100$nm at a height $%
h<h_{\max }$ to make an effective protection. Luckily, this drawback 
can be overcome.  In Sec. III we propose a method to increase the 
thickness of He film by using the surface roughness. Our estimates in 
Sec. III show the feasibility of a strong reduction of UCN losses due to the 
absorption inside trap walls. 

3. Liquid He introduces new scattering mechanisms of UCN: the He vapor atoms
and the soft thermal excitations -- the quanta of surface waves, called
ripplons. The corresponding scattering rates were recently calculated for a
neutron on the lowest vertical level \cite{PhysRevC.94.025504}, and these
estimates are qualitatively valid for higher levels. The energy of $^{4}$He
evaporation is $7.17$K per atom. Hence, the concentration of vapor
exponentially decreases with temperature, $n_{V}\propto \exp \left( -7.17/T%
\left[ K\right] \right) $. The scattering on He vapor can be disregarded at
low $T<0.5K$. However, such a low temperature is very essential; at twice
higher $T=1K$ the neutron lifetime due to scattering rate on He vapor is
only $2.3$ min. The ripplons, the quanta of surface waves, are gapless
excitations with soft spectrum: $\omega _{q}^{2}=\left( \sigma _{He}/\rho
_{He}\right) \left( q^{2}+a_{H}^{-2}\right) q$. The neutron scattering rate $%
w_{R}$ on ripplons depends linearly on temperature \cite{PhysRevC.94.025504}
and, formally, cannot be discarded even below $0.5K$. Luckily, the amplitude
of neutron-ripplon interaction is small, and the corresponding neutron
scattering time $1/w_{R}$ exceeds many hours at $T<0.5K$.  
Moreover, the linear temperature dependence of neutron-ripplon scattering 
rate $w_{R}\left( T\right) $ allows it to be effectively taken into account 
by extrapolation to zero temperature. 
Nevertheless, to avoid new scattering mechanisms of UCN, introduced 
by helium, low temperature $T<0.5K$ is required.

4. The liquid $^4$He covering the trap walls must be isotopically pure, because $^3$He absorbs neutrons. 
The $^4$He isotope purification technology was developed long ago \cite{JEWELL1982,HENDRY1987,Hayden2006}. 
The superfluid heat flush method allows reducing the $^3$He concentration to less than 
$5\times 10^{-13}$ \cite{HENDRY1987}, which is sufficient to neglect the UCN losses due to the 
absorption by $^3$He. Note, that this degree of isotopic purification is achieved even in a continuous 
flow apparatus at a production rate of 3.3 m$^3$h$^{-1}$ at STP \cite{HENDRY1987}.

\emph{II. Advantages.}

1. For large neutron traps of size $L\gg h_{\max }$ the most important
neutron losses are due to their collisions with the trap bottom. This is
clear by just comparing the corresponding surface areas: their ratio is $%
\sim h_{\max }/L$. Liquid He covers the trap bottom as a thick film,
sufficient for complete protection of neutrons with energy $E<V_{0}^{He}$
from being absorbed inside the trap material.

2. As shown in Sec. III, due to the combination of capillary effects with
the van der Waals attraction, the surface roughness of trap walls strongly
increases the effective He film thickness on them. For various roughness
configurations, e.g., as discussed in Sec. III, one can easily achieve He
film thickness on the walls to be sufficient for almost complete UCN
protection from absorption inside trap material. This may strongly, by
1-2 orders of magnitude, diminish the neutron losses inside material traps. 

3. Superfluid He fills cavities and other surface roughness, thus
removing their negative effect on the neutron storage time in material
traps. The negative effect of surface roughness without He was estimated to
increase 2-3 times the neutron losses from wall absorption\cite%
{Ignatovich/1990}. Eliminating this negative effect is a considerable
advantage, which also leads to longer UCN storage times.

\medskip

The above list and discussion of weak and strong points of liquid He film
covering UCN traps shows that the potential advantage is much stronger
than the drawbacks. A possible experiment may use similar methods as 
in current UCN $\tau_n$ measurements with walls covered by Fomblin oil 
\cite{Serebrov2008PhysRevC.78.035505,Serebrov2018PhysRevC.97.055503}. 
Such an experiment contains five stages \cite{Serebrov2008PhysRevC.78.035505,Serebrov2018PhysRevC.97.055503}: 
(1) filling, (2) monitoring (with spectrum preparation), (3) holding, 
(4) empting, and (5) measurement of background.
Helium film may prevent the filling of a UCN trap with neutrons using 
standard valves in the trap bottom, but the neutrons can be filled from the top 
either (1) with the help of an additional time-dependent potential created by a
magnetic field, or (2) by moving up and/or rotating the trap itself, 
as in Refs. \cite{Serebrov2008PhysRevC.78.035505,Serebrov2018PhysRevC.97.055503}. 
Since the gravitational potential difference is only $100$ neV/m, 
and the magnetic field creates a potential $60$ neV/T, a small vertical 
magnetic-field gradient $16$mT/cm compensates for the rise of UCN kinetic 
energy when they are filled from the top of the trap for half of UCN 
with one spin projection. A specially designed time-dependent magnetic 
field may even decelerate and move the neutrons \cite{PhysRevA.72.043619}, of course,
conserving their phase-space density according to the Liouville's theorem. 

The accuracy of proposed $\tau_n$ measurements using such UCN traps covered by liquid 
helium is limited by (1) the statistical error, (2) uncertainties in 
the rate of inelastic neutron scattering by excitations of liquid helium, and 
(3) the remaining UCN absorption in the trap wall material, if some part of 
the wall is covered by too thin He film. The statistical error depends very much 
on the experimental setup and procedure. It can be reduced only by 
increasing the UCN density in the holding stage or by performing several 
cycles/replicas of $\tau_n$ measurement. 
The inelastic scattering rate $1/\tau_i$ 
of UCN by surface and bulk excitations in liquid helium was estimated 
\cite{PhysRevC.94.025504} to be rather small at $T<0.5$K, when the concentration 
of He vapor is exponentially small. The main contribution to UCN scattering rate  
at $T<0.5$K comes from ripplons -- the quanta of surface waves. For a neutron on the lowest 
vertical quantum level but with large in-plane neutron energy $K=100$neV, 
the rate $1/\tau_{r}$ of its scattering to continuous spectrum via ripplon 
absorption was roughly upper
estimated \cite{PhysRevC.94.025504} to be less than $1/\tau^{up}_{r}\approx 2\times 
10^{-5}$s$^{-1}\times T$[K] (see Eq. (D17) of Ref. \cite{PhysRevC.94.025504}). 
However, the UCN scattering rate in our problem is considerably less 
than the above estimate $1/\tau^{up}_{r}$ for several reasons. 
The first and, probably, most important difference between our case and the one in 
Ref. \cite{PhysRevC.94.025504} is that in Ref. \cite{PhysRevC.94.025504} the neutron 
was on the lowest energy level in the vertical direction. Hence, the typical value 
$\psi_{\perp} (0)$ of its normalized wave function on the helium surface $z=0$ 
was rather large, because this wave function extends to only a distance 
$L_{\psi}\sim 10\mu$m from the surface. In our case the neutron wave function is 
localized on a distance $L_{\psi}\leq h_{\max} =18$cm above the He surface, which is $10^4$ times larger. 
Hence, the normalization coefficient of this wave function, entering $\psi_{\perp} (0)$, 
is $100$ times smaller. Since the UCN scattering rate by ripplons is
$1/\tau_r\propto \vert \psi_{\perp} (0)\vert ^2 $, this may reduce by orders of 
magnitude the rate of ripplon absorption by UCN with typical out-of-plane component 
of kinetic energy $E_{\perp}\sim V_0^{He}/5$.   
 Second, the in-plane neutron velocity $v_{||}$ in our problem is much smaller than in Ref. \cite{PhysRevC.94.025504} 
and does not exceed $v_{||\max}=\sqrt{2V_0^{He}/m}=1.9$m/s. Since the scattering rate $\tau_{r}^{-1}$ 
increases with the increase of $v_{||\max}$, this factor also reduces $\tau_{r}^{-1}$ in our case. 
Third, the calculations in Ref. \cite{PhysRevC.94.025504} are too rough, 
because they are aimed to show that $\tau_{r}^{-1}$ is negligibly small for a different experiment.    
Our current preliminary estimates even for the lowest vertical neutron level, 
i.e., neglecting the first factor above, give that  
$1/\tau_{r} < 10^{-5}$s$^{-1}\times T$[K]. At $T=0.5$K this $\tau^{up}_{r}$ corresponds 
to UCN loss probability $<0.4$\% of $\beta$-decay probability, which 
is already much better than in the most precise UCN $\tau_n$ measurements
and can be easily accounted for using the temperature extrapolation of 
measured $\tau_n(T)$ to $T=0$ due to its linear $T$-dependence.   
Note that the predicted $\tau^{-1}_{r}(T)\propto T$ with very high precision, 
because the typical ripplons, important for UCN scattering, have energy 
$\hbar\omega_r \sim V_0^{He}\ll T$, hence, their equilibrium density $\propto T$. 
The problem of inelastic UCN scattering is also important 
in current $\tau_n$ measurements using Fomblin oil 
\cite{MAMBOPhysRevLett.63.593,PICHLMAIER2010,Serebrov2017,ArzumanovPhysLettB2015,Serebrov2008PhysRevC.78.035505,Serebrov2018PhysRevC.97.055503}.
Even with a nonlinear temperature dependence of inelastic scattering rate 
in these experiments the standard temperature extrapolation allows reducing 
the systematic error due to inelastic neutron scattering 
by an order of magnitude. The linear temperature dependence of the ripplon 
absorption rate $\tau^{-1}_{r}$ allows it to be accounted for with much higher 
precision by similar temperature extrapolation of measured $\tau_n(T)$ to $T=0$. 
Thus, if helium temperature is reduced to $T=0.2$K, the 
expected systematic error of UCN $\tau_n$ value due to ripplon scattering is less than $0.01$\%. 
More accurate calculations of the UCN scattering rate by ripplons, similar to those in 
Ref. \cite{PhysRevC.94.025504} but taking the velocity distribution of UCN in our particular problem, 
most probably, will further reduce the estimate of UCN scattering rate by ripplons.   
The last important factor, affecting the accuracy of $\tau_n$ measurements, 
is the remaining UCN absorption in the trap wall material, if some parts of 
the wall are covered by too thin He film. Although this error is greatly reduced 
by using liquid He, it may still be present due to any local distortions of wall roughness. 
This type of errors can be effectively taken into account by the standard size extrapolation method, 
similar to that in Refs. \cite{MAMBOPhysRevLett.63.593,Serebrov2008PhysRevC.78.035505,Serebrov2018PhysRevC.97.055503}, 
which makes the corresponding systematic error negligibly small in our case.           

To summarize, liquid He film may reduce a hundred times the
neutron absorption rate inside UCN trap material. This superfluid He film can be made thicker 
than the neutron penetration depth due to capillary effects by using a rough 
wall surface. This has a potential to strongly improve the
accuracy of neutron lifetime measurements and of other experiments with UCN, 
important for particle physics, astrophysics, and cosmology. It may also help 
to explain the discrepancy between measured neutron $\beta $%
-decay times using cold neutron beams and UCN magnetic and material traps.

\begin{acknowledgments}

P.G. acknowledges the support of the Ministry of Science and Higher
Education of the Russian Federation in the framework of Increase Competitiveness Program of NUST ''MISiS'' Grant No. K2-2020-038. A.D.
acknowledges the State Assignment No. 0033-2019-0001.

\end{acknowledgments}

\appendix


\section{Calculation of the neutron wave function near the
He-covered trap wall}

First consider the trap wall covered by helium film, as shown in Fig. \ref%
{FigScheme}a.

\subsection{Sewing of wave functions at $x=0$}

The boundary conditions on the neutron wave function at $x=0$ are written as%
\cite{LL3}%
\begin{equation}
\psi \left( -0\right) =\psi \left( +0\right) , ~ \psi ^{\prime }\left(
-0\right) =\psi ^{\prime }\left( +0\right) .  \label{Boundaryx0}
\end{equation}%
Substituting the wave functions (\ref{psi1}) and (\ref{psi2}) to Eq. (\ref%
{Boundaryx0}) gives%
\begin{equation}
A=B_{1}-B_{2}, ~ A\kappa _{W}=\kappa _{He}\left( B_{1}+B_{2}\right) ,
\end{equation}%
or 
\begin{equation}
B_{2}=B_{1}\frac{\kappa _{W}-\kappa _{He}}{\kappa _{W}+\kappa _{He}}, ~ A=%
\frac{B_{1}2\kappa _{He}}{\kappa _{W}+\kappa _{He}}.
\end{equation}%
These relations can be rewritten as%
\begin{equation}
B_{1}=A\frac{\kappa _{W}+\kappa _{He}}{2\kappa _{He}}, ~ B_{2}=A\frac{\kappa
_{W}-\kappa _{He}}{2\kappa _{He}}.
\end{equation}

Let us relate the coefficients $B_{1}$, $B_{2}$, and $B$ in Eq. (\ref%
{psi2}). Since 
\begin{gather}
\psi _{II}\left( x\right) =B\sinh \left( \kappa _{He}\left[ x+x_{0}\right]
\right) = \\
=\frac{B}{2}\exp \left( \kappa _{He}\left[ x+x_{0}\right] \right) -\frac{B}{2%
}\exp \left( -\kappa _{He}\left[ x+x_{0}\right] \right) ,
\end{gather}%
we obtain%
\begin{equation}
B_{1}=\frac{B}{2}\exp \left( \kappa _{He}x_{0}\right) , ~ B_{2}=\frac{B}{2}%
\exp \left( -\kappa _{He}x_{0}\right) ,
\end{equation}%
and%
\begin{equation}
B=2\sqrt{B_{1}B_{2}}=2B_{1}\sqrt{\frac{\kappa _{W}-\kappa _{He}}{\kappa
_{W}+\kappa _{He}}}=A\sqrt{\frac{\kappa _{W}^{2}}{\kappa _{He}^{2}}-1}.~
\label{B}
\end{equation}%
The coordinate shift 
\begin{equation}
x_{0}=\frac{1}{2\kappa _{He}}\ln \left( \frac{B_{1}}{B_{2}}\right) =\frac{1}{%
2\kappa _{He}}\ln \left( \frac{\kappa _{W}+\kappa _{He}}{\kappa _{W}-\kappa
_{He}}\right) .  \label{x0}
\end{equation}%
Usually, $\kappa _{He}/\kappa _{W}<\kappa _{0He}/\kappa _{0W}\ll 1$; if the
wall material is beryllium, $\kappa _{0He}/\kappa _{0W}\approx 0.27$. Then 
\begin{equation}
B\approx A\kappa _{W}/\kappa _{He}, ~ x_{0}\approx 1/\kappa _{W}.
\label{ABApp}
\end{equation}

\subsection{Sewing at $x=d_{He}$}

The boundary conditions for the wave functions (\ref{psi2}) and (\ref{psi3})
at the point $x=d_{He}$ are written as 
\begin{equation}
\psi \left( d_{He}-0\right) =\psi \left( d_{He}+0\right) ;~\psi ^{\prime
}\left( d_{He}-0\right) =\psi ^{\prime }\left( d_{He}+0\right)
\label{BoundarydHe}
\end{equation}%
and give 
\begin{eqnarray}
B\sinh \left[ \kappa _{He}\left( d_{He}+x_{0}\right) \right] &=&C\sin \left[
k\left( d_{He}+x_{1}\right) \right] , \ \ \ \ \  \label{Eq1} \\
B\kappa _{He}\cosh \left[ \kappa _{He}\left( d_{He}+x_{0}\right) \right]
&=&Ck\cos \left[ k\left( d_{He}+x_{1}\right) \right] . \ \ \ \ \ \   \label{Eq2}
\end{eqnarray}%
One can relate the coefficients $B$ and $C$ without calculating $x_{1}$,
using the trigonometric identity%
\begin{eqnarray}
1 &=&\left( B\sinh \left[ \kappa _{He}\left( d_{He}+x_{0}\right) \right]
/C\right) ^{2} + \label{CEq} \\
&&+\left( B\kappa _{He}\cosh \left[ \kappa _{He}\left( d_{He}+x_{0}\right) %
\right] /Ck\right) ^{2},  \notag
\end{eqnarray}%
which gives%
\begin{eqnarray}
C &=&B\left\{ \left( \sinh \left[ \kappa _{He}\left( d_{He}+x_{0}\right) %
\right] \right) ^{2}\right. +  \label{CB} \\
&&+\left. \left( \cosh \left[ \kappa _{He}\left( d_{He}+x_{0}\right) \right]
\kappa _{He}/k\right) ^{2}\right\} ^{1/2}.  \notag
\end{eqnarray}%
Using Eqs. (\ref{B}) and (\ref{CB}), we obtain Eq. (\ref{CA}).

The shift $x_{1}$ can be found from the ratio of Eqs. (\ref{Eq1}) and (\ref%
{Eq2}):%
\begin{equation}
k\tanh \left[ \kappa _{He}\left( d_{He}+x_{0}\right) \right] =\kappa
_{He}\tan \left[ k\left( d_{He}+x_{1}\right) \right] ,
\end{equation}%
which gives Eq. (\ref{x1}).


%

\end{document}